\documentclass[showpacs,preprintnumbers,amsmath,amssymb]{revtex4}
\usepackage{graphicx}
\begin{document}
\preprint{IMAFF-RCA-06-08}
\title{Some notes on the Big Trip}

\author{Pedro F. Gonz\'{a}lez-D\'{\i}az}
\affiliation{Colina de los Chopos, Centro de F\'{\i}sica ``Miguel A.
Catal\'{a}n'', Instituto de Matem\'{a}ticas y F\'{\i}sica Fundamental,\\
Consejo Superior de Investigaciones Cient\'{\i}ficas, Serrano 121,
28006 Madrid (SPAIN).}
\date{\today}
\begin{abstract}
The big trip is a cosmological process thought to occur in the
future by which the entire universe would be engulfed inside a
gigantic wormhole and might travel through it along space and time.
In this paper we discuss different arguments that have been raised
against the viability of that process, reaching the conclusions that
the process can actually occur by accretion of phantom energy onto
the wormholes and that it is stable and might occur in the global
context of a multiverse model. We finally argue that the big trip
does not contradict any holographic bounds on entropy and
information.
\end{abstract}


\maketitle


\pagebreak

Rather bizarre implications from dark energy models are now being
considered that might ultimately make the future of the universe
sings a somehow weird melody. It has been in fact recently proposed
[1] that if the current value of the equation-of-state parameter $w$
would keep up being less than -1 in the future, then the throat
radius of naturally existing wormholes could grow large enough to
engulf the entire universe itself, before this reached the so called
big rip singularity [2], at least for an asymptotic observer. This
rather astonishing result - which has been dubbed the "big trip" -
has proved to be not free from a number of difficulties which has
been raised afterwards and that mainly includes: (1) The result is
obtained by using a static metric and therefore it has been claimed
[3] that the accretion of dark energy cannot significantly change
the amount of exotic matter in the wormhole and hence no large
increase of the throat radius should be expected; (2) wormhole
space-times are all asymptotically flat and thereby a very large
increase of the throat size would imply that the insertion of the
wormhole cannot be made onto our universe [4]; (3) quantum
catastrophic creation of vacuum particles on the chronology horizon
would make macroscopic wormholes completely unstable [5,6], and (4)
the holographic bound on the entropy [7,8] would prevent any
relevant amount of information to flow through the wormholes, so
that these wormholes could never be used to circumvent the big rip
singularity [9]. The present report aims at discussing these four
difficulties. It will be seen that none of the problems (1), (3) and
(4) indeed hold for the asymptotic observer, and that problem (2) is
a debatable one and might require considering the big trip to take
place within the context of a multiverse scenario.

\vspace*{0.5cm}

\noindent {\bf 1.} Metric staticity may not be a real problem
actually. The question that has been posed is that by using a static
metric one automatically ensures that there cannot be any energy
flux of the exotic stuff making the wormhole and therefore no
arbitrarily rapid accretion can take place, let alone the accretion
of the entire universe required by the big trip. It has been in this
way claimed that a static metric may well justify small accretion
rates by rigorously calculating the integrated stress tensor
conservation laws [10] needed to suitably evaluating the accretion
of dark energy according to the generalized Michel theory developed
by Babichev, Dokuchaev and Eroshenko [11], but it can never describe
extreme accretion regimes. However, we shall show in what follows
that by using the static four-dimensional Morris-Thorne metric [12]
with a zero shift function we obtain exactly the same result on such
extreme regimes (that is a big trip) as when we introduce any time
dependence in the $g_{rr}$ metric tensor component entering that
metric. Let us start with the static Morris-Thorne metric [12] with
zero shift function,
\begin{equation}
ds^2= -dt^2 +\frac{dr^2}{1-\frac{K(r)}{r}} +r^2
d\Omega_2^2.\nonumber
\end{equation}
From just the integration of the conservation laws for the
momentum-energy tensor and its projection on four-velocity (energy
flux), the following two equations can then be obtained [10]:
\begin{equation}
\frac{ur^2 e^{\int_{\rho_{\infty}}^{\rho}d\rho/(p+\rho)}}{m^2
\sqrt{1-\frac{K(r)}{r}}}=A
\end{equation}
\begin{equation}
\frac{\sqrt{\frac{u^2+\frac{K(r)}{r}}{1-\frac{K(r)}{r}}}}{e^{
\int_{\rho_{\infty}}^{\rho}d\rho/(p+\rho)}}(p+\rho)
=B=\hat{A}\left(\rho_{\infty}+p(\rho_{\infty})\right),
\end{equation}
where $m$ is the exotic mass which can be assumed to be spherically
distributed on the wormhole throat, $u=dr/ds$, $K(r)$ is the shape
function [12], $A$, $B$ and $\hat{A}$ are generally positive
constants, and the dark-energy pressure, $p$, and energy density,
$\rho$, bear all time-dependence in these two expressions. Moreover,
since Eq. (1) [where the constant $A$ must be dimensionless (note
that we are using natural units so that $G=c=\hbar=1$)] should
describe a flux of dark energy onto the wormhole, $u>0$ and energy
conservation ought then to imply that the exotic mass of the
wormhole -and hence the radius of its throat- should progressively
change with time, as a consequence from the incoming dark energy
flux. Thus, even though the starting metric is static, the energy
stored in the wormhole must change with time by virtue of dark
energy accretion, in the model considered in Refs. [1] and [10].

Now, in order to implement the effect of a dark energy flux onto the
wormhole implied by Eq. (1), we must introduce a general rate of
change of the energy stored in the wormhole due to the external
accretion of dark energy onto the wormhole. From the momentum
density it can be derived that the rate of change of the exotic mass
is generally given by $\dot{m}=\int dST^r_0$, where $T^{\mu}_{\nu}$
is the dark momentum-energy tensor of the universe containing a
Morris-Thorne wormhole, and $dS=r^2 \sin\theta d\theta d\phi$.
Hence, using a perfect-fluid expression for that momentum-energy
tensor, $T_{\mu\nu}=(P+\rho)u_{\mu}u_{\nu}+g_{\mu\nu}p$, and Eqs.
(1) and (2), we obtain finally for $p=w\rho(t)$,
\begin{equation}
\dot{m}=-4\pi m^2 A\hat{A}\sqrt{1-\frac{K(r)}{r}}(p+\rho) ,\nonumber
\end{equation}
For the relevant asymptotic regime $r\rightarrow\infty$, the rate
$\dot{m}$  reduces to $\dot{m}=-4\pi m^2 A\hat{A}(1+w)\rho(t)$,
whose trivial integration using the general phantom scale factor
$a(t)=a_0(1-\beta(t-t_0))^{-2/[3(|w|-1)]}$, with $\beta$ a positive
constant, yields an increasing expression for $m(t)$ leading to the
big trip, provided $w<-1$ and $r\rightarrow\infty$, that is [1,10]
\begin{equation}
m\propto K_0=\frac{K_{0i}}{1-\frac{4\pi Q
K_{0i}(|w|-1)(t-t_0)}{(1-\beta(t-t_0))}} ,\nonumber
\end{equation}
where $K_{0i}$ is the initial value of the radius of the wormhole
throat and $Q$ is a positive constant. Mere inspection of this
equation tells us that the big trip ($K_0\rightarrow\infty$) takes
place quite before than big rip ($a\rightarrow\infty$) does.

Note that for $r<\infty$ there will be no big trip, but just an
increase of the size of the wormhole throat that ceases to occur at
a given time. However, because the metric is static, the exotic
energy-momentum tensor component describing any internal radial
energy flux $\Theta^r_0=0$, so that, even though the dark
energy-momentum tensor component $T^r_0\neq 0$, it has been claimed
that accretion of phantom energy following this pattern could only
be valid for small rates $\dot{m}$, so that at first sight such a
mechanism could not describe arbitrary dark-energy accretion rates
and even less so a regime in which the entire universe is accreted.

A more careful consideration leads nevertheless to the conclusion
that a big trip keeping the same characteristics as those derived
from a static metric stands up as a real phenomenon even when we use
a non static metric in such a way that $\Theta^r_0\neq 0$ and
$T^r_0\neq 0$ simultaneously. This result can be seen to be a
consequence from the fact that the big trip can only occur
asymptotically, at $r\rightarrow \infty$. In fact, if we start with
the corresponding, simplest time-dependent wormhole metric with zero
shift function,
\begin{equation}
ds^2= -dt^2 +\frac{dr^2}{1-\frac{K(r,t)}{r}} +r^2
d\Omega_2^2,\nonumber
\end{equation}
(where the shape function $K(r,t)$ is allowed to depend on time both
when tidal forces are taken into account and when such forces are
disregarded. In the latter case, all time-dependence in the metric
is concentrated on the radius of the wormhole throat, that is it is
assumed that the wormhole evolves with time by changing its overall
size, while preserving its shape, such as it is thought to occur
during the big trip) then Eqs. (1) and (2) would be modified to read
\begin{equation}
\frac{ur^2 e^{\int_{\rho_{\infty}}^{\rho}d\rho/(p+\rho)}}{m^2
\sqrt{1-\frac{K(r,t)}{r}}}=AD^{1/2} \nonumber
\end{equation}
\begin{eqnarray}
&&e^{-\int_{\rho_{\infty}}^{\rho}\frac{d\rho}{p+\rho}}D^{1/2}\left\{T_0^r
+\frac{\dot{K}_0 K_0}{r}\left[\frac{E}{r^2}(p-T_r^r)
+\frac{F(T_0^0-T_r^r)}{r^2}\right.\right.\nonumber\\
&&\left.\left.\frac{D^{1/2}}{r}\int
dr\left(rED^{1/2}\left(\frac{2F(p-T_r^r)}{r^5 D^2}
+\frac{d(p-T_r^r)}{r^2 D dr}\right) +\frac{Fd(T_0^0-T_r^r)}{rD
dr}\right)\right]\right\} = B ,\nonumber
\end{eqnarray}
where $A$ and $B=\hat{A}\left(\rho_{\infty}+p(\rho_{\infty})\right)$
are again constants and
\begin{equation}
D=1-\frac{K(r,t)}{r}\nonumber
\end{equation}
\begin{equation}
E=\frac{2}{3}(r^2+2K_0^2)\nonumber
\end{equation}
\begin{equation}
F=r^2-2K_0^2 .\nonumber
\end{equation}
In these expressions we have particularized, for the aim of
simplicity, in the case of a wormhole with zero tidal forces. The
first of them adds an extra factor $D^{1/2}=\sqrt{1-K(r,t)/r}$ to
the constant $A$ of Eq. (1) and the second one contains an extra
term depending on $\dot{K}_0 K_0$ and an overall extra factor
$D^{1/2}$, respect to Eq. (2). It is worth noticing that from the
field equations associated with the above time-dependent solution we
have $\dot{K}(r,t)=-8\pi (r-K(r,t))^2\Theta_0^r$. Now, while
$\dot{K}(r,t)=0$ would imply $\Theta_0^r=0$ and hence a vanishing
radial flux of the exotic stuff making the wormhole, this does not
imply that there is no incoming radial energy flux from the dark
energy of the universe. It can then be readily seen that whereas the
extra term and factors involved at the precedent two expressions
would remarkably change the mass rate equation for any $r<\infty$,
the extra term vanishes and the factors become unity in the
asymptotic regime, $r\rightarrow\infty$, where the big trip would be
expected to take place, so that these expressions reduce to Eqs. (1)
and (2) asymptotically and hence the relevant mass rate equation
keeps up being $\dot{m}=-4\pi m^2 A\hat{A}(1+w)\rho(t)$ on that
regime where we therefore recover the big trip feature even for the
above time-dependent metric. There thus exists at least a particular
example of a simple time-dependent wormhole metric which leads to a
big trip when the wormhole accretes phantom energy. Of course, using
the most general initial metric
$ds^2=-e^{\Psi(r,t)}dt^2+e^{\Phi(r,t)}+R(r,t)d\Omega_2^2$, one
cannot generically show that accretion of phantom energy leads to a
big trip.

On the other hand, the inflationary effect of the universal speed-up
on wormhole metric has already been considered in the case that
there is no accretion of dark energy [13]. It gives a time-dependent
metric which is comoving with the background, such as it was again
obtained in Ref. [3]. Although that result is no doubt correct, it
has nothing to do with the accretion of dark energy onto the
wormhole, which is the case considered in Ref. [1]. In fact, a
recent calculation of the accretion of dark energy onto black holes
leading even to the vanishing of the black holes at the big rip
singularity for $w<-1$ also uses a static metric, that is either the
Schwarzschild metric [11] or the Kerr metric [14], without employing
any exact solution for the black hole in a cosmological space-time
that shows a time evolution displaying such an extreme behavior.
Thus, the evaluation of accretion energy onto wormholes carried out
in Refs. [1] and [10] is not only applicable to a regime of low
accretion rate but to any unboundedly large accretion rates and
therefore the results obtained in these references are also correct.

Even so, if anyone insisted in having a metric describing by itself
a wormhole in a Friedmann universe with time evolution induced by
accretion of phantom energy, displaying the big trip feature, it can
be seen that such a metric may still be actually built up by simply
inserting a suitable dimensionless factor $W^2$ depending on the
scale factor $a(t)$, into the three-dimensional spatial part of e.g.
the Morris-Thorne metric [12], i.e. generically
\[ds^2=-e^{2\Phi(r)}dt^2 +W(t)^2\left(\frac{dr^2}{1-\frac{K(r)}{r}}
+r^2 d\Omega_2^2\right) ,\] with
\[W(t)=\frac{1}{1- \frac{K_0}{\kappa}\left[\left(\frac{a}{a_0}\right)^{3(|w|-1)/2}-1\right]}
,\] where $\kappa$ is a positive constant. The factor $W(t)$
consistently becomes unity when either the radius of the throat
$K_0\rightarrow 0$ (i.e. when the wormhole pinches off) or $a=a_0$,
with $a_0$ the initial value of the scale factor. One could expect
that this metric is really the exact solution that corresponds to a
given distribution of exotic matter whose total amount varies with
time at exactly the rate described in Ref. [1] due to phantom energy
accretion. This can be readily seen by using the notation
$\exp(-\lambda(r,t))= W(t)^2 (1-K/r)^{-1}$, so that
$\dot{\lambda}=-2\dot{W}W$; hence and from the field equations we
can then have $d(W^2)/dt=-8\pi r(1-K/r)t_0^r$, with $t_{\mu\nu}$ the
energy-momentum tensor for this case. Inserting the above expression
for $W(t)$ and taking again for the scale factor of the universe
$a(t)=a_0(1-\beta(t-t_0))^{-2/[3(|w|-1)]}$, with $\beta$ a positive
constant, we finally get that, although the energy flux $t_0^r$
vanishes at the big rip, it does not so at the necessarily previous
time when the trip occurs [1], so implying that phantom energy
accretion takes place all the way from $t=t_0$ up to the big trip.

Moreover, by considering (i) a proper circumference $c$ at the
wormhole throat, $r=K=K_0$, $\theta=\pi/2$, at any constant time, we
get in the accelerating framework $c=K_0\int_0^{2\pi}d\phi W(t)=2\pi
K_0W(t)$, and (ii) a radial proper length through the wormhole
between any two points $A$ and $B$ at constant time, which is given
by $d(t)=\pm W(t)\left(\sqrt{r_B^2-K_0^2}-\sqrt{r_B^2-K_0^2}\right)$
for $K(r)=K_0^2/r$, it can be easily seen that the above solution
displays a big trip, as it can also be readily shown that the form
of that metric is preserved with time [13,15] and the factor $W$
blows up when the scale factor reaches the critical value
$a=a_{bt}\equiv
a_0\left(1+\frac{\kappa}{K_0}\right)^{2/[3(|w|-1)]}$, which takes
place before the occurrence of the big rip, such as it happens in
the case studied in Ref. [1].

A comment is worth mentioning at this point. Whereas the accretion
mechanism used above is classical in nature the very structure of
the wormholes should be quantum mechanically considered on some
regimes [16]. In particular, sub-microscopic wormholes have been
shown to be stabilized by quantum mechanical effects that induce a
possible discretization of time [17]. We note however that our
approximation can safely be applied to wormhole sizes which widely
separate from those where quantum effects are expected to be
important. Moreover, being true that both the energy density and
curvature of the universe increase with time, it is easy to check
that these quantities only acquires the sufficiently high values
approaching the Planck scale that requires a proper quantization of
space-time [18] as one comes close to the big rip singularity, a
regime still far enough from that characterizing the big trip
phenomenon as to allow one to take the classical approach to be
reliable. In fact, at the time when the wormhole throat starts
exceeding the radius of the universe, the value of the scale factor,
and hence of the energy density and curvature of the universe are
expected to be many orders of magnitude smaller than their
counterparts at the close neighborhood of the big rip.

\vspace*{0.5cm}

\noindent {\bf 2.} Difficulty (2) is a debatable one. It states that
if wormholes are asymptotically flat and their throat is allowed to
grow larger than the universe itself, then the insertions of the
grown up wormholes could by no means be kept on the universe where
the wormholes were originally formed. The implication of this is
twofold. On one hand, the dark-energy accretion process by which
wormholes grow up larger than their mother-universe refers only to
an asymptotic observer [10] and could therefore occur in the context
of a multiverse [19], where the grown-up wormholes would re-insert
in other universes which are larger than the mother-universe. Though
it does not appear quite clear how that re-insertion can be
implemented, the extension implies in turn the creation of physical
connections among the universes. In addition, asymptotic flatness by
itself is not a problem; if the matching conditions between the
wormhole and the cosmological metric can be satisfied, then there is
no reason, in principle, why an otherwise asymptotically flat
wormhole cannot be matched to a universe.

On the other hand, the involvement of other universes and the lack
of any common time concept for the components of the
wormhole-coupled universes would fully eliminate any problems
related to causality violation in the global context of the
multiverse during the re-insertion process and therefore, although
the universe that originally nested the wormholes could actually
time travel and thereby avoid the big rip singularity relative to
its local framework, the whole system would not actually undergo any
time travel or causality violation. Thus, even though the so-called
"big trip" may mean a disruption of the causal evolution of our
universe in its local future, the consideration of a multiverse
scenario actually leads to the preservation of causality in the
global framework of the multiverse and, relative to it, the "big
trip" term would rather refer to an information transfer process
between two of its constituting universes. At the very least, that
information transfer between two large universes could be viewed to
eventually provide a proof for the existence of the multiverse
itself and a formal suitable starting point to construct a quantum
field theory for cosmology.

\vspace*{0.5cm}

\noindent {\bf 3.} Quantum instability of wormholes could indeed be
a real problem to preserve the existence of these space-time
tunnelings in a phantom universe, at least on the regions
sufficiently far from the big rip singularity where the big trip may
happen. The catastrophic quantum creation of particles on the
chronology (Cauchy) horizon would be expected to wipe off any trace
of macroscopic wormholes evolving at times quite before that of the
big rip if the throat of these wormholes would grow at a rate
smaller than or nearly the same as the speed of light. However,
since the throat growing rate induced by phantom energy accretion
clearly exceeds the speed of light asymptotically, particles created
by the quantum excitation of vacuum would never reach such
chronology horizon and the wormhole would keep asymptotically stable
even quantum-mechanically. This would be the second explicit example
of a violation of the Hawking's chronology protection conjecture.
The first one corresponded to the self-consistent vacuum and was
derived by Gott and Li [20]. The present violation would actually
extend to any topological generalizations of the flat Misner space
[21], other than wormholes, and leave in this way an open door for
the big trip to take place in the future. This will be now
explicitly considered by using a two-dimensional ($\theta,\phi =$
constant) version of the Morris-Thorne metric [12] with zero shift
function, $\Phi(r)=0$, which can be written as
\begin{equation}
ds^2= -dt^2+\frac{dr^2}{1-K(r)/r} .
\end{equation}
For the asymptotic region where the big trip takes place, the
simplest wormhole metric with $K(r)=K_0^2/r$ (which corresponds to a
wormhole with zero tidal forces and where $K_0$ is the radius of the
wormhole throat) becomes flat and we can therefore convert it into a
Rindler-like metric, with $t=\xi\sinh\eta$, $r=\xi\cosh\eta$, so
that
\begin{equation}
ds^2=-\xi^2 d\eta^2 +d\xi^2 ,
\end{equation}
which just covers the right quadrant of the Minkowski space, i.e.
the region $r > |t|$. The reflection $(\eta,\xi)\rightarrow
(\eta,-\xi)$ would lead to the description of the left quadrant of
Minkowski space, i.e. the region $r < - |t|$. A metric like that
given by Eq. (4) can also be derived if we keep up $K_0^2/r^2$
generally nonzero and introduce the definitions
\begin{equation}
t=\xi\sinh\eta ,\;\;\; \sqrt{r^2-K_0^2}=\xi\cosh\eta .
\end{equation}
In this case, metric (4) covers the region $\sqrt{r^2-K_0^2}>|t|$
and the left quadrant is also obtained by the above reflection and
describes the region with $\sqrt{r^2-K_0^2}<-|t|$. We shall consider
in what follows the more general case where $K_0^2/r^2$ is taken to
be generally nonzero.

Now a Misner-like space can be obtained by identifying points so
that the Misner symmetry
\begin{equation}
\left(t,\sqrt{r^2-K_0^2})\rightarrow (t\cosh(nb)+
\sqrt{r^2-K_0^2}\sinh(nb), t\sinh(nb)+
\sqrt{r^2-K_0^2}\cosh(nb)\right),
\end{equation}
in which $n$ is an integer number and $b$ a boost constant, be
satisfied. Under such an identification the points $(\eta,\xi)$ in
$R$ (or $L$) are identified with the points $(\eta+nb,\xi)$ in $R$
(or $L$), so that regions $R$ and $L$ contains timelike curves
(CTCs). The Cauchy horizons (which in this case coincide with the
chronology horizons) that separate the above two regions are placed
at $\sqrt{r^2-K_0^2}=\pm t$, i.e. at $\xi\rightarrow 0$. We note
that in the general case $r=K_0$ on the Cauchy horizons, at least
for $\eta$ finite. Thus, the Cauchy horizon becomes the same as the
the apparent horizon which makes the metric (1) singular. This is a
coordinate singularity rather than a real curvature singularity as
it can be seen by introducing an extension of the metric where such
a singularity is no longer present. In fact, by introducing the
advanced and retarded coordinates, $u=t-\sqrt{r^2-K_0^2}$,
$v=t+\sqrt{r^2-K_0^2}$ , we obtain $ds^2=-dudv$, which can be seen
to show no singularity at $r=K_0$.

Using a Rindler self-consistent vacuum [20], one can now derive the
Hadamard two-point function for a conformally coupling scalar field
for our two-dimensional Misner-like space to be [20]
\begin{equation}
G^{(1)}(X,X')=\sum_{n=-\infty}^{\infty}\frac{1}{2\pi^2}\times
\frac{\gamma}{\xi\xi ' \sinh\gamma\left[-(\eta-\eta ' +nb)^2
+\gamma^2\right]},
\end{equation}
where the parameter $\gamma$ is in the present case defined by
$\cosh\gamma=(\xi^2 +\xi'^2)/(2\xi\xi')$. Now, with the usual
definition of the Hadamard function for the Minkowski vacuum
$G_M^{(1)}$ reduced to our two-dimensional case, one can derive the
regularized Hadamard function $G^{(1)}_{reg} =G^{(1)}-G_M^{(1)}$,
and hence the renormalized stress-energy tensor in region $R$. This
turns out to become proportional to $[(2\pi/b)^4 -1]/\xi^4$. It is
thus seen that even though for a space that had $r^2\leq t^2+K_0^2$,
and hence $\xi^2\leq 0$, and $b\neq 2\pi$, the renormalized
stress-energy tensor would diverge at the Cauchy horizon $\xi=0$, if
either $b=2\pi$ or we had $r^2 > t^2+K_0^2$ (i.e. if we confine the
system to be inside the $R$ quadrant without touching its
boundaries), i.e. $\xi^2
> 0$, then this tensor would be convergent everywhere in regi\'{o}n $R$,
because the expression derived above for it can be in this case
valid only for $\xi>0$ (recall that for the big trip
$r\rightarrow\infty$ while $t$ and $K_0$ are both finite when the
size of the wormhole throat overtakes that of the universe) which
does not include the Cauchy horizon which never is fully well
defined. It could be said that the particles created by vacuum
polarization cannot reach the chronology horizon unless at the
moment at which $K_0$ becomes infinite; i.e. when the wormhole
ceases to exist relative to any observers. The first of these two
stabilized situations ($b=2\pi$) corresponds to the Li-Gott
self-consistent vacuum [20] and the second one should be associated
with the case where we had a big trip, for which
$r\rightarrow\infty$.

The moderation or possible removal of the future singularity that
can be induced by quantum effect of matter [22] and quantum gravity
[23] appear again to be not significantly influencing the regime
before the big trip where the semi-classical approximation used in
this section applies. That approximation would therefore remain as a
sufficiently accurate procedure. After the big trip and on the
regime approaching the big rip, wormholes simply cease to exist [1].
On the other hand, it can be stressed that the existence of a future
singularity has nothing to do with the emergence of the big trip
phenomenon.

\vspace*{0.5cm}

\noindent {\bf 4.} We shall finally consider in a little more detail
the last of the above-alluded difficulties, that is the one related
with the holographic bound on the entropy. Let us be then a little
more explicit mathematically. The entropy($S$)-energy($E$) bound
which was introduced by Bekenstein [7] for a spherical system of
radius $B$ can be written as
\begin{equation}
S\leq\frac{2\pi EB}{\hbar c} .
\end{equation}
This inequality implies [7] a limit in the information that can be
drawn from the given system such that
\begin{equation}
I<\frac{2\pi EB}{\hbar c \log 2} .
\end{equation}
The holographic bound for entropy can be derived from the above
entropy-energy bound and reads [8]
\begin{equation}
S<\frac{A}{4\ell_P^2} ,
\end{equation}
where $A$ is the surface area and $\ell_P$ is the Planck length.
These bounds must all apply to the entropy and information which are
allowed to traverse one of the growing wormhole during the big trip.
Thus, $S$ refers to the entropy of the observable matter that
traverses the grown-up wormhole and has therefore nothing to do with
the phantom stuff. Hence, the rather peculiar properties of phantom
thermodynamics [24,25] do not influence at all the analysis to
follow.

Taking now for the radius $B$ the proper value of the uppermost
horizon
\begin{equation}
B=a(t)\int_0^{t_{max}}\frac{dt'}{a(t')} ,
\end{equation}
in which $a$ is the scale factor of the universe, from Eq. (9) we
obtain then for reasonable values of the involved cosmological
parameters [9] and $t_{max}=t_*$, with $t_*$ the time at which the
big rip takes place, in the approximation of full phantom energy
dominance, i.e.
\[a(t)=a_0\left(1-\frac{t-t_0}{t_*-t_0}\right)^{-\frac{2}{3(|w|-1)}},\]
(where $t_* =t_0+\frac{2}{3(|w|-1)\sqrt{8\pi \rho_0/3}}$ and $a_0$,
$t_0$ and $\rho_0$ are the initial values of the scale factor, time
and energy density, respectively), that $I<I_{max}\simeq 100 $ bits.
This figure is certainly very small and, by far, does not allow the
components of any future advanced civilization to make any big trip.
This is difficulty (4). However, the upper integration limit
$t_{max}=t_*$ in Eq. (11) is only strictly valid if the big rip
singularity cannot be avoided by the action of local, smaller
wormholes branched off from the region in the close neighborhood of
that singularity. Wormholes able to connect the regions before and
after the big rip (note that for $t>t_*$ the value of the scale
factor $a$ becomes decreasing with time $t$ and keeps up real and
positive for an infinite family of discrete values of $w$),
short-cutting the singularity, have been in fact shown [25] to
copiously crop up on that neighborhood and become stabilized by
cosmological complementarity. Such wormholes can allow a flux of
unboundedly large information carried by real signaling or matter in
both directions. The action of these wormholes would then physically
extend the evolution of the universe up to an infinite time. If so
then we have $t_{max}\rightarrow\infty$ and the use of the bound (9)
leads to the result that the maximum involved information that can
be transferred during the big trip process should be infinite, so
ultimately allowing the universe itself and any future advanced
civilizations to be transferred as a whole by means of such a
process. Thus, the holographic and entropy-energy bounds do not
preclude the trip of our own universe through gigantic, stable
wormholes grown up by accretion of phantom energy.

The big trip appears then a real possibility to occur in the future
of the universe if the equation-of-state parameter of this would
preserve a value less than -1 long enough in the future. Of course,
all of what has been discussed in this paper has a rather
speculative character and is based on the conception that the
observed accelerating expansion of the universe is due to the
presence of a quintessence scalar field that behaved like a
dark-energy fluid. Since other models could also be invoked to
justify the observations, the big rip and its implications would
remain as just an interesting possibility.

\acknowledgements

\noindent This work was supported by MCYT under Research Project No.
BMF2002-03758. The author benefited from discussions with C.
Sig\"{u}enza, A. Rozas and S. Robles.


\begin{references}
\bibitem {4} There is no topologically allowed way by which an asymptotically flat
wormhole tunneling with the throat larger than a given sphere can be
inserted into the interior of that sphere.
\bibitem {15} Using the embedding of a $t$=const, $\theta=\pi/2$
slice of the time-dependent metric in a flat three-dimensional
Euclidean space with cylindrical metric, $ds^2=d\bar{z}^2+d\bar{r}^2
+\bar{r}^2 d\phi^2$, we can get $\bar{r}=W(t)r|_{t=const}$,
$d\bar{r}^2=W(t)^2 dr^2|_{t=const}$. Now, relative to the
($\bar{z},\bar{r},\phi$) coordinate system, the form of the wormhole
metric will be preserved provided that the metric on the embedded
slice has the form $ds^2= (1-\bar{K}(\bar{r})/\bar{r})d\bar{r}^2
+\bar{r}^2 d\phi^2$, with $\bar{K}(\bar{r})$ having a minimum at
some $\bar{K}_0 =\bar{r}_0$, which is a condition that is easily
fulfilled [13].
\end{references}
\end{document}